\documentclass[10pt,twocolumn,showpacs,amsmath,amssymb,floatfix,superscriptaddress]{revtex4-2}

\usepackage{graphicx}
\usepackage{dcolumn}
\usepackage{bm}
\usepackage{color}
\usepackage{txfonts}
\usepackage{microtype}
\usepackage{hyperref}
\usepackage[english]{babel}
\usepackage{slashed}
\usepackage{gensymb}
\usepackage{epsfig}
\usepackage[normalem]{ulem}
\usepackage{amsmath}
\usepackage{array}
\usepackage{booktabs}
\allowdisplaybreaks[4]

\addto\captionsenglish{}
\begin{document}
\title{Ultrafast Spin Rotation of Relativistic Lepton Beams via Terahertz Wave in a Dielectric-Lined Waveguide }

\author{Zhong-Peng Li}
\affiliation{Ministry of Education Key Laboratory for Nonequilibrium Synthesis and Modulation of Condensed Matter, Shaanxi Province Key Laboratory of Quantum Information and Quantum Optoelectronic Devices, School of Physics, Xi'an Jiaotong University, Xi'an 710049, China}

\author{Yu Wang}
\affiliation{Ministry of Education Key Laboratory for Nonequilibrium Synthesis and Modulation of Condensed Matter, Shaanxi Province Key Laboratory of Quantum Information and Quantum Optoelectronic Devices, School of Physics, Xi'an Jiaotong University, Xi'an 710049, China}

\author{Ting Sun}
\affiliation{Ministry of Education Key Laboratory for Nonequilibrium Synthesis and Modulation of Condensed Matter, Shaanxi Province Key Laboratory of Quantum Information and Quantum Optoelectronic Devices, School of Physics, Xi'an Jiaotong University, Xi'an 710049, China}

\author{Feng Wan}
\affiliation{Ministry of Education Key Laboratory for Nonequilibrium Synthesis and Modulation of Condensed Matter, Shaanxi Province Key Laboratory of Quantum Information and Quantum Optoelectronic Devices, School of Physics, Xi'an Jiaotong University, Xi'an 710049, China}

\author{Yousef I. Salamin}
\affiliation{Department of Physics, American University of Sharjah, Sharjah, POB 26666 Sharjah,  United Arab Emirates}

\author{Mamutjan Ababekri}
\affiliation{Ministry of Education Key Laboratory for Nonequilibrium Synthesis and Modulation of Condensed Matter, Shaanxi Province Key Laboratory of Quantum Information and Quantum Optoelectronic Devices, School of Physics, Xi'an Jiaotong University, Xi'an 710049, China}

\author{Qian Zhao}
\affiliation{Ministry of Education Key Laboratory for Nonequilibrium Synthesis and Modulation of Condensed Matter, Shaanxi Province Key Laboratory of Quantum Information and Quantum Optoelectronic Devices, School of Physics, Xi'an Jiaotong University, Xi'an 710049, China}

\author{Kun Xue}
\affiliation{Ministry of Education Key Laboratory for Nonequilibrium Synthesis and Modulation of Condensed Matter, Shaanxi Province Key Laboratory of Quantum Information and Quantum Optoelectronic Devices, School of Physics, Xi'an Jiaotong University, Xi'an 710049, China}

\author{Ye Tian}
\affiliation{State Key Laboratory of High Field Laser Physics and CAS Center for Excellence in Ultra-intense Laser Science, Shanghai Institute of Optics and Fine Mechanics, Chinese Academy of Sciences, Shanghai, People’s Republic of China}
\affiliation{Center of Materials Science and Optoelectronics Engineering, University of Chinese Academy of Sciences, Beijing, People’s Republic of China}

\author{Wen-Qing Wei}
\affiliation{Ministry of Education Key Laboratory for Nonequilibrium Synthesis and Modulation of Condensed Matter, Shaanxi Province Key Laboratory of Quantum Information and Quantum Optoelectronic Devices, School of Physics, Xi'an Jiaotong University, Xi'an 710049, China}

\author{Jian-Xing Li}\email{jianxing@xjtu.edu.cn}
\affiliation{Ministry of Education Key Laboratory for Nonequilibrium Synthesis and Modulation of Condensed Matter, Shaanxi Province Key Laboratory of Quantum Information and Quantum Optoelectronic Devices, School of Physics, Xi'an Jiaotong University, Xi'an 710049, China}
\affiliation{Department of Nuclear Physics, China Institute of Atomic Energy, P. O. Box 275(7), Beijing 102413, China}

\date{\today}
\begin{abstract}
	
Spin rotation is central for the spin-manipulation of lepton beams which, in turn, plays an important role in investigation of the properties of spin-polarized lepton beams and the examination of spin-dependent interactions. However, realization of compact and ultrafast spin rotation of lepton beams, between longitudinal and transverse polarizations, still faces significant challenges. Here, we put forward a novel method for ultrafast (picosecond-timescale) spin rotation of a relativistic lepton beam via employing a moderate-intensity terahertz (THz) wave in a dielectric-lined waveguide (DWL). The lepton beam undergoes spin precession induced by the THz magnetic field. We find that optimizing the lepton velocity and THz phase velocity in the DLW can mitigate the impact of transverse Lorentz forces on the lepton beam and increase the precession frequency, thereby maintaining the beam quality and enhancing the efficiency of transverse-to-longitudinal spin rotation. The final polarization degree of the lepton beam exceeds $98\%$, and the energy spread can be improved significantly. Flexibility in adjusting the electromagnetic modes within the DLW adds further potential for spin-manipulation, and holds promise for advancing the development of spin-polarized particle beams, which have broad applications in materials science and atomic, nuclear, and high-energy physics.
	
\end{abstract}

\maketitle

Beams of relativistic spin-polarized (SP) leptons are finding important applications in materials science \cite{Gay09AAMOP} and atomic~\cite{Danielson15RMP, kessler85polarized}, nuclear  \cite{Aidala13RMP}, and high-energy physics \cite{Moortgat08PR,Vauth16IJMPCS}, etc. For instance, longitudinally spin-polarized (LSP) lepton beams play a profound role in detection of the spin structures of nucleons~\cite{Anderle21FP,Aidala13RMP} and atoms \cite{Danielson15RMP}, exploring magnetic properties of materials and molecules \cite{Gay09AAMOP}, and probing the dynamics of ultrafast magnetic fields in extreme environments \cite{GZ21PRL,GZ23PRL}. Furthermore, beams of the transversely spin-polarized (TSP) variety have many applications in high-energy physics, ranging from the Higgs particle production \cite{Rindani09PRD} to CP asymmetries \cite{Bartl07JHEP} in the search for new physics beyond the Standard Model. 

The main process underlying the generation of relativistic SP lepton beams is based on the Sokolov-Ternov effect \cite{Sokolov66}, which relies on the spin-dependent radiation in storage rings and produces TSP lepton beams. In addition, with the rapid development of ultraintense ultrafast laser techniques, it has become possible to achieve laser pulse peak intensities in excess of $\rm 10^{23}W/cm^2$ \cite{Mourou85OC, Yoon21OPtica, Kawanaka16JPCS, Cartlidge18S, Danson19HPLSE, Bahk04OL, Tiwari19OL, Pirozhkov17OE, Guo18OE, Zou15HPLSE, Gales18RPP, Bromage19HPLSE, Yaron18S}, which opens up new avenues for the production of ultrafast polarized lepton beams \cite{Xue23PRL, Piazza12RMP, Fedotov23PR, Song22PRL, Salamin06PR, Dai22MRE, Xue22FR}. To that end, various schemes, based on asymmetric laser fields and helicity transfer, have been proposed, including nonlinear Compton scattering using elliptically polarized \cite{Wan20PLB,LYF19PRL} or bichromatic laser fields~\cite{Seipt19PRA,CYY19PRL}, as well as the linear \cite{Zhao22PRD} and nonlinear \cite{Vranic18SR, Xie17MRE} Breit-Wheeler processes of colliding high-energy $\gamma$-photons. At present, however, these schemes can only generate TSP lepton beams. Although LSP beams can be generated using the Bethe-Heitler process, high repetition is required due to the relatively low charge \cite{Variola14NIMA}, and the average degree of polarization being in the range of 30\%-40\% \cite{Abbott16PRL}. 

Considering the fact that only transverse spin-polarization can prevent depolarization in storage rings \cite{Malysheva12IPAC}, and production of highly polarized, high-charge LSP lepton beams faces constraints, well-known spin rotators \cite{Moffeit06} or Siberian snakes \cite{Derbenev78PA} are required at conventional accelerators \cite{Mane05JPNPP, Huang18PRL, Moortgat08PR, Nikitin19CEPC}. These devices ensure preservation of the polarization of the lepton beam during propagation while enabling the selection of the desired polarization direction at the interaction point. Traditional spin rotators rely on dipole magnets and superconducting solenoids. 
The length of these spin rotators is typically on the order of tens of meters \cite{WWQ23PRRes}. Once installed, the spin rotator, optimized for a specific particle energy, may require substantial reconfiguration, or even redesign, to accommodate beams of different energies, resulting in significant challenges in terms of facility design, cost, limited tunability, and operational flexibility.


\begin{figure}[t]
	\vspace{6pt}
	\centering\includegraphics[width=1\linewidth]{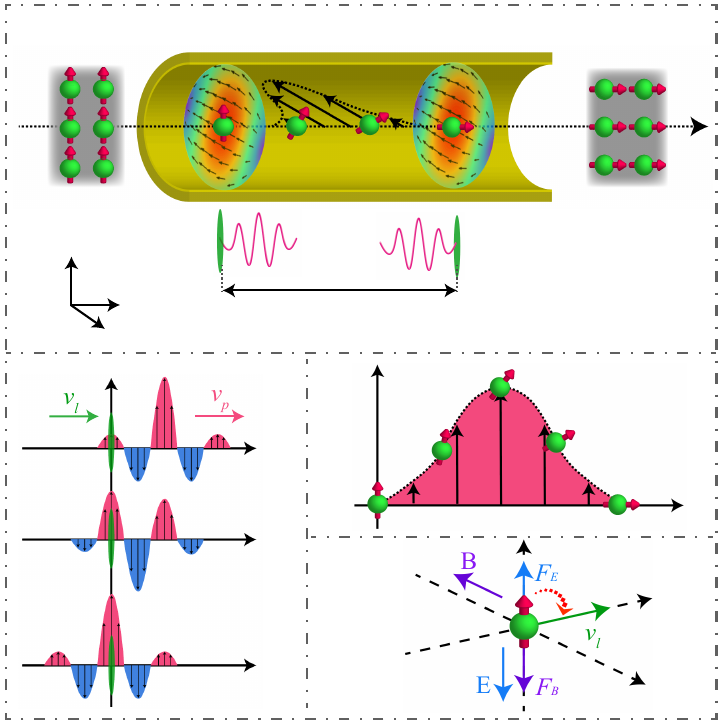}
	\begin{picture}(300,0)
		\put(3,247){ (a)}
		\put(56,245){$\rm TE_{11}$ mode}
		\put(120,225){B}
		\put(8,185){TSP lepton} \put(17,175){bunch}
		\put(197,185){LSP lepton} \put(206,175){bunch}
		\put(82,150){$L=\tau v_g v_l/(v_l-v_g)$}
		\put(27,167){$x$}
		\put(42,154){$z$}
		\put(37,144){$y$}
		\put(4,128){(b)}
		\put(60,50){phase}
		\put(60,40){matching}
		\put(85,96){$\varphi$}
		\put(85,65){$\varphi$}
		\put(85,22){$\varphi$}
		\put(29,126){B}
		\put(107,128){(c)}
		\put(232,78){$t$}
		\put(118,108){\rotatebox{90}{B}}
        \put(192,128){spin-}
        \put(192,118){manipulation}
		\put(107,63){(d)}
		\put(182,68){$x$}
		\put(222,20){$y$}
		\put(220,43){$z$}

	\end{picture}
	\setlength{\abovecaptionskip}{-0.5 cm}
	\caption{Scenarios of lepton spin-manipulation using a THz wave in a DLW. (a) A TSP lepton beam is converted into a LSP beam, where the red arrows indicate the spin vectors. The $\rm TE_{11}$ mode and magnetic field envelope seen by leptons is displayed in the DLW. (b) Spatiotemporal correlation between the lepton beam and THz magnetic field profiles in phase space. Green ellipses represent lepton beams, and pink and light-blue parts represent the magnetic field oscillations. (c) Variation with time of the magnetic field experienced by leptons during spin-manipulation. (d) Directions of the electric field (light-blue arrow) and magnetic field (purple arrow), along with the corresponding forces they exert on the lepton. Green arrow represents lepton velocity, and red arcuate arrow indicates spin precession direction.}
	\label{fig1}
\end{figure}

\begin{figure}[!t]
	\centering
		\includegraphics[width=1\linewidth]{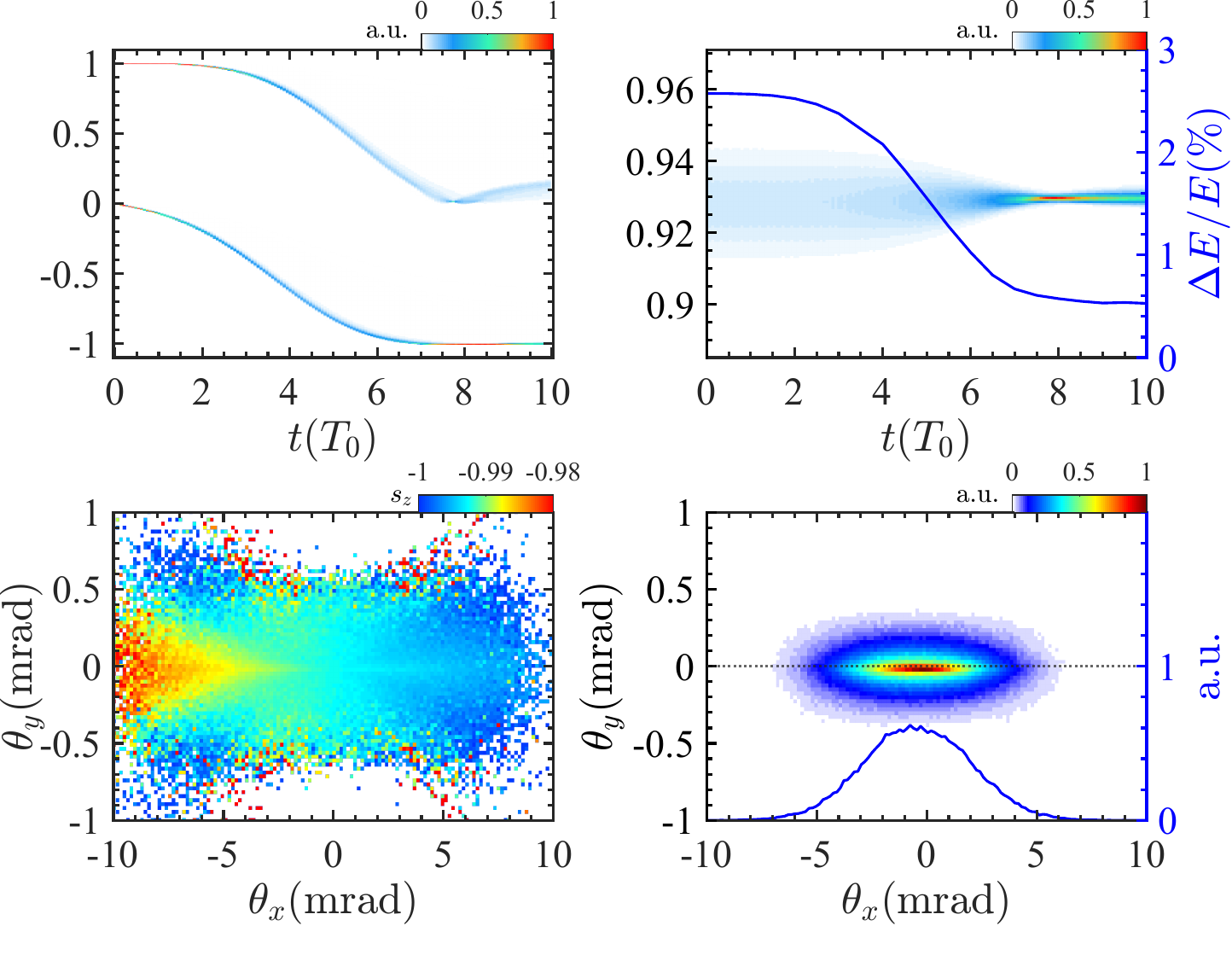}
		
		\begin{picture}(300,0) 
			\put(25,182){(a)}
			\put(0,155){\rotatebox{90}{$s$}}
			\put(40,145){$s_z$}
			\put(75,175){$s_\perp$}
			
			\put(143,182){(b)}
			\put(115,145){\rotatebox{90}{$\varepsilon_e(\rm MeV)$}}
%
			\put(22,91){ (c)}
			
			\put(141,90){ (d)}
		\end{picture}
	\setlength{\abovecaptionskip}{-0.9 cm}
	\caption{(a) Variation, with time $t$ and spin-polarization $s$, of normalized number density of the electron beam. $T_0$ is period of the THz wave. (b) Variation,  with time $t$ and electron energy $\varepsilon_e$, of normalized number density (pcolor), and temporal evolution of energy spread $\Delta E/E$ (solid line). (c) Angle-resolved distributions in $(\theta_x,\theta_y)-$space of longitudinal spin-polarization of the electron beam, after passing through a DLW. The transverse deflection angles are defined as $\theta_x = \arctan (p_x/p_z)$ and $\theta_y = \arctan (p_y/p_z)$, respectively. (d) Number density of the electron beam (pcolor) in $(\theta_x,\theta_y)-$space and the corresponding beam profile (solid line) after exiting the DLW. Intensity of incident THz wave corresponds to $a_0=0.3$ ($E_0\sim1.61\times10^9$ V/m for $\lambda = 0.6 \rm ~mm$) and its duration is $\rm 4.5~ ps$. Initial electron energy is $\varepsilon_0=0.93$ MeV, and phase and group velocities of the THz wave are $v_p = 0.935c$ and $v_g = 0.72c$, respectively.}
	\label{figsr}
\end{figure}

\begin{figure}[!t]
%
	
		\centering
	    \includegraphics[width=1\linewidth]{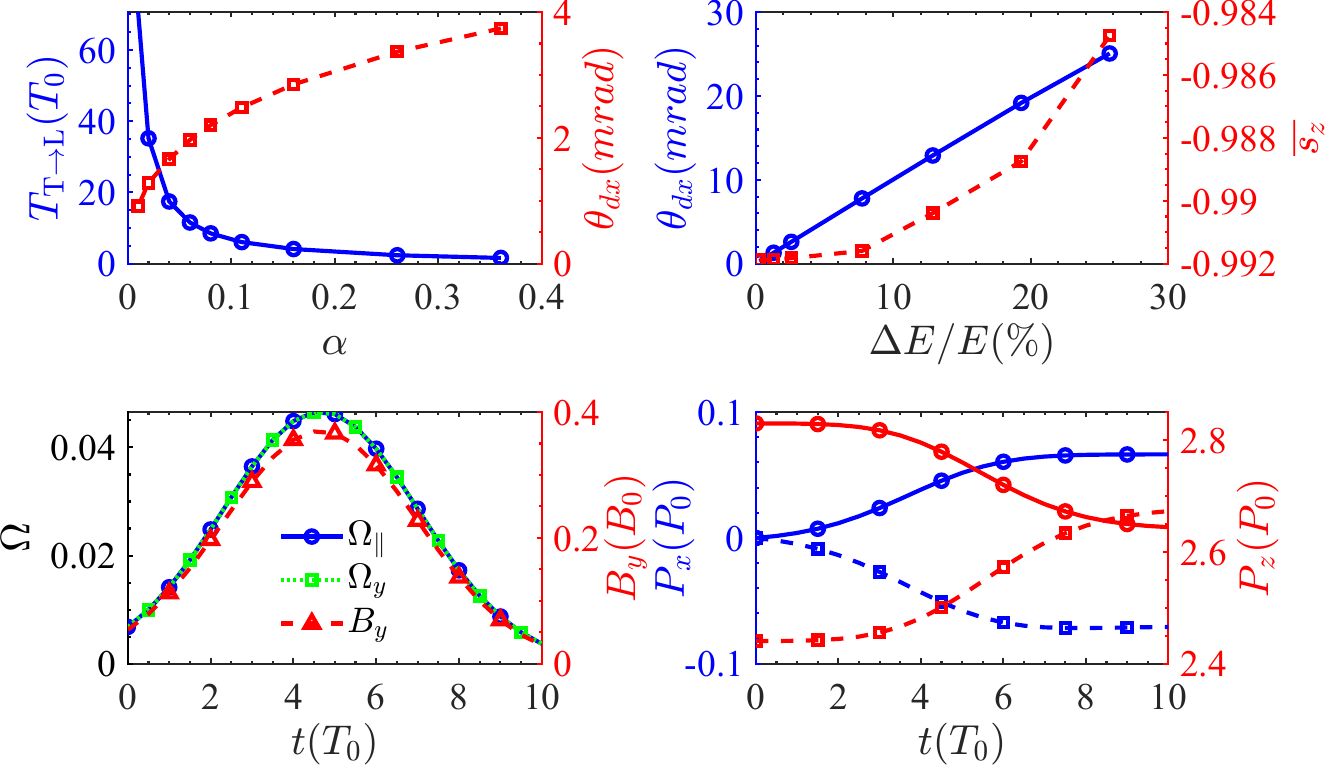} \vspace{25pt}
	    \begin{picture}(300,0) 
	    	\put(25,143){(a)}
	    	\put(141,143){(b)}
	    	
	    	\put(25,68){(c)}

	    	\put(141,68){(d)}

	    \end{picture}

	     \label{minipage1}
    \vspace{-20pt}
%
%
%

%
%
   \setlength{\abovecaptionskip}{-0.4 cm}
    \caption{(a) Variations, with $\alpha$, of the time $ T_{\rm T\to L}$ required to rotate from TSP to LSP (solid line) and the rms divergence along the $x-$axis $\theta_{dx}$ (dashed line). Energy spread of  initial electron beam is  $2.6\%$, everywhere. (b) Variations, with the initial energy spread, of $\theta_{dx}$ (solid line) and average spin-polarization of the electron beam exiting the DLW (dashed line). Mean energy of electron beam is 0.93~MeV. (c) Temporal evolution of $y$-axis precession frequency $\Omega_y$, parallel precession frequency $\Omega_\parallel$, and average intensity of magnetic field $B_y$ experienced by electrons. Precession frequency and magnetic field intensity are normalized to $\omega$ and $B_0=m_e \omega c/e$, respectively. (d) Temporal evolution of transverse and longitudinal momenta (red and blue lines, respectively) for electrons at energies $1.1\varepsilon_0$ (solid lines) and $0.9\varepsilon_0$ (dashed lines). Electron momentum is normalized to $P_0=m_ec$. }
    \label{fig3}

\end{figure}

In this Letter, we explore the possibility of utilizing a terahertz (THz) wave in a dielectric-lined waveguide (DLW)~\cite{Nanni15NV,Xu21NPho,Fisher22NPho,ZDF20PRX,Tang21PRL,Wong13OE,YXQ23NPho,Pacey19PRAB} to attain spin rotation capability. A phase-matched $\rm TE_{11}$ mode~\cite{Fisher22NPho} in an optimized DLW can enable the conversion of a TSP lepton beam \cite{Hibberd20NPho} to a LSP beam at the picosecond timescale; see Fig.~\ref{fig1}(a) for a schematic. By aligning the initial spin-polarization of the lepton beam perpendicular to the magnetic field and matching the lepton velocity $v_l$ to the THz phase velocity $v_p$, we enable effective spin rotation from TSP to LSP through spin precession induced by the THz magnetic field. Duration of the lepton beam is about tens of femtoseconds \cite{Kuwahara14APL, ST24PRL}, significantly shorter than the THz period. Thus, utilization of a THz wave is advantageous for achieving spin-manipulation of the whole lepton beam. Combined with the proximity of $v_l$ to $v_p$, we can ensure that the lepton beam stays in its optimal spin-manipulation phase while propagating through the DLW; see Fig.~\ref{fig1}(b). The optimal phase can be achieved by tuning the phase delay between the THz wave and the leptons, which can be governed precisely by adjusting the carrier envelope phase (CEP) of the THz wave~\cite{Kawada16OL}. Besides, the phase matching preserves the direction of the magnetic field experienced by the leptons, and variation of the spin-manipulation strength is related to the shape of the THz pulse envelope; see Fig.~\ref{fig1}(c). Owing to co-propagation of the lepton beam and the THz wave, the transverse electric force $F_E$ and magnetic force $F_B$ exerted on the leptons act in opposite directions. When $v_l\approx v_p$, the values of $F_E$ and $F_B$ become comparable~\cite{supp}, leading to their mutual cancellation; see Fig.~\ref{fig1}(d). This suppresses transverse deflection of the lepton beam. Unlike electromagnetic (EM) waves propagating in vacuum at the speed of light $c$, a DLW effectively governs the phase velocity $v_p$ and the group velocity $v_g$ of the THz waves, providing ideal conditions for spin-manipulation. The final spin-polarization exceeds $98\%$, and the beam energy spread can be notably improved; see Fig.~\ref{figsr}. We find that $v_l$ and $v_p$ significantly affect the spin-manipulation efficiency; see Fig.~\ref{fig3}. Our method is feasible using currently available THz sources, and can achieve spin-manipulation over a picosecond time-scale, with high efficiency.

In our simulations, we use a beam of electrons as a representative example of leptons. With an eye on the damage threshold of DLW \cite{Wang89IEEE,Thompson08PRL,Hassanein06PRL} and the currently available energies to which electrons can be accelerated by THz-driven accelerators \cite{Tang21PRL,YXQ23NPho}, we employ THz wave intensities corresponding to $0.1\lesssim a_0 \equiv eE/(m_e\omega c) \lesssim 1$ and an electron energy at the MeV level. Here, $E$, $\omega$, $m_e$, and $e$ are the electric field amplitude and angular frequency of the incident THz wave,  and the mass and charge of the electron, respectively. Our scheme is feasible, on account of the fact that the maximum energy and peak power of currently available THz pulses can reach about 55 mJ \cite{LGQ19PNAS} and $\rm 1 ~TW$~\cite{LGQ20PRX}, respectively. Given these values, the quantum nonlinear parameter~\cite{Gonoskov22RMP,Baier98,Ritus85JSLR}, which characterizes the quantum radiation effects, is $\chi=[e\hbar/(m_e^3 c^4)]\sqrt{-(F_{\mu\nu}p^\nu)^2}\sim10^{-8}\ll 1$, where $\hbar$, $F_{\mu\nu}$, and $p^\nu$ are the reduced Planck's constant, field tensor and four-momentum of the particle, respectively. Electron velocity is denoted by $v_e = c\beta_e$, where $\beta_e$ is the electron velocity normalized to $c$. Besides, the Stern-Gerlach and space-charge forces are much weaker than the Lorentz force, for the parameters of our simulations~\cite{Mane05RPP,Thomas20PRAB}. Thus, the influence of radiation on the electron spin is negligible, and the dynamics of the particle's spin and momentum are mainly governed by the Thomas-Bargmann-Michel-Telegdi (T-BMT)~\cite{1999Jackson} and relativistic Lorentz equations \cite{05Paul}, respectively. Moreover, longitudinal size of the electron beam is considerably smaller than the THz wavelength and the propagation distance is relatively short. Thus, all electrons are manipulated at approximately the same phase, and the wakefields resulting from the beam's EM fields reflecting off the DLW walls are negligible \cite{Wong13OE,Kim10PRSTAB}. The DLW structure considered here consists of a vacuum core with a single  inner dielectric layer and outer copper coating. To achieve effective spin-manipulation, the group and phase velocities of the THz wave propagating in the DLW can be adjusted by varying thickness and type of the dielectric layer and inner diameter of the DLW \cite{Tang21PRL,Wong13OE,ZDF20PRX}. The initial phase experienced by the electrons can be controlled by adjusting the CEP of the THz wave, using a CEP shifter \cite{Kawada16OL}.  

A THz wave in a DLW offers distinct advantages, compared to the infrared laser-driven dielectric microstructure \cite{Peralta13N,Breuer13PRL,England14RMP,Freemire23PRAB}, such as intense field strength, accessible fabrication and high beam-charge \cite{Hibberd20NPho}. By contrast to the conventional means of employing a transverse magnetic (TM) mode for lepton acceleration, the combination of a cylindrical DLW with a phase-velocity-matched transverse electric (TE) mode can manipulate the lepton spin-polarization more efficiently and accurately. On the one hand, TE modes have little influence on the longitudinal velocity of the leptons due to lack of a longitudinal electric field. On the other hand, TE modes exhibit weaker surface electric fields than TM modes, allowing the DLW to accommodate higher internal EM field strengths \cite{Wang89IEEE,Thompson08PRL,Hassanein06PRL}. 

\textbf{Spin rotation:} The $\rm TE_{11}$ mode, which can be excited by a linearly polarized THz wave \cite{Gallot00JOSA}, exhibits a relatively uniform EM field distribution at the center of the DLW. This homogeneity guarantees that electrons undergo consistent spin precession across the transverse plane. By taking advantage of the homogeneous EM field and the synchronized electron velocity, we can achieve efficient and controllable spin rotation, enabling the conversion of TSP beams to LSP beams. From Fig. \ref{figsr}(a) we can see that, under the manipulation of the THz in DLW, a TSP electron beam can be converted into a LSP beam with the spin-polarization exceeding 98\%. The final spin distribution is highly concentrated, as can be seen in \cite{supp}. The reverse conversion from LSP to TSP can also be achieved with equal efficiency~\cite{supp}. Due to the initial energy spread in the beam, some of the electrons' velocities may not be completely aligned with $v_p$. Electrons with velocities higher than the THz phase velocity experience transverse acceleration followed by longitudinal deceleration, influenced by the transverse EM field. Conversely, electrons with lower velocities experience longitudinal acceleration. Thus, implementation of the $\rm TE_{11}$ mode also has the capability to significantly decrease the beam's energy spread. The energy spread is lowered from 2.6\% to 0.5\%, as shown in Fig. \ref{figsr}(b). The initial energy spread can also lead to an expansion of the root-mean-square (rms) divergence of the electron beam. By compressing initial energy spread of the electron beam, the rms divergence can be effectively reduced \cite{supp}. Besides, the final average degree of polarization also depends on the initial angle between the spin vector and the magnetic field \cite{supp}. In order to achieve efficient conversion between LSP and TSP beams, it is crucial to ensure that the energy spread is low and that initial spin-polarization of the electron beam is perpendicular to magnetic field of the THz wave. 

\textbf{Manipulation of spin-polarization:} This primarily involves alteration of the longitudinal spin-polarization. The rate of change of longitudinal polarization can be written as
$\frac{d}{dt}\left(\hat{\bm{\beta_e}}\cdot\bm{s}\right) =-(e/m_ec)s_\perp \alpha B$ \cite{1999Jackson,supp}.
Here, $\alpha = \left(g/2-1\right)-\left(g\beta_e/2-1/\beta_e \right)v_p/c \in (0,\infty)$ is the longitudinal precession frequency coefficient, related to the precession frequency of the longitudinal spin component $ \Omega_{\parallel}=-(e/m_ec) \alpha B$. Also, $g$, $\bm{s}$, $s_\perp$ and $\hat{\bm{\beta_e}}$ are the Land$\rm \acute{e}$ $g-$factor, electron spin, spin component perpendicular to the electron velocity and a unit vector in the direction of $\bm{\beta_e}$, respectively. In the specific case of spin rotation from TSP to LSP with $v_e \approx v_p$, $\alpha \approx g(1-\beta_e^2)/2 \in (0,g/2)$. Note that it is difficult to rotate the electron spin in vacuum using an oscillating plane-wave EM field, because the effects on the spin due to two successive half-cycles cancel each other out, unless an asymmetric laser is used \cite{WWQ23PRRes}. Fortunately, $v_p$ and $v_g$ can be controlled flexibly in a DLW, by altering type and thickness of the dielectric in the DLW and varying frequency of the THz waves \cite{Tang21PRL,ZDF20PRX}. By controlling $v_p$, we can ensure that direction of the EM field experienced by electrons in a DLW remains unchanged.

The longitudinal precession frequency coefficient $\alpha$ plays a crucial role in the design of spin-rotation experiments, since it characterizes the spin rotation efficiency of the electrons. We investigate effect of $\alpha$ on the spin rotation efficiency when a plane THz wave propagates in the DLW. It can be seen from Fig. \ref{fig3}(a) that $T_{\rm T \to L}$, the time required to rotate from TSP to LSP, is inversely proportional to $\alpha$. Furthermore, we can infer that our method can enable ultrafast spin-manipulation. Using a THz wave of wavelength 0.6 mm and intensity $a_0=0.3$, as an example, it takes merely 12 ps to achieve ultrafast spin rotation from TSP to LSP. Increasing the THz intensity can further shorten the spin rotation time. With velocity dependence of the transverse Lorentz forces exerted on electrons in mind, we investigate influence of the initial energy spread on the quality of the electron beam. From Fig. \ref{fig3}(b), it is evident that the initial energy spread affects the rms divergence quite significantly. Taking into account the transverse distribution of spin-polarization, density of the electron beam \cite{supp} and the final average spin-polarization (see Fig. \ref{fig3}(b)), it is recommended to keep the initial energy spread below $5\%$.

In our simulations, electric field direction of the $\rm TE_{11}$ mode aligns with the $x$-axis, and the magnetic field extends along the $y$-axis, with no longitudinal magnetic field present. Our estimates indicate that the precession frequencies around the $x-$ and $z-$axes, namely $\Omega_x$ and $\Omega_z$, are roughly $10^{-8}$, or negligibly small compared to $\Omega_y$. Hence, $\Omega_y$ effectively represents the overall precession frequency. As can be seen from Fig.~\ref{fig3}(c), the longitudinal precession frequency is in agreement with $\Omega_y$, suggesting that within the $\rm TE_{11}$ mode, the calculation of precession frequency of electrons can be performed utilizing $\alpha$ and the magnetic field. In addition, envelope of the magnetic field that electrons encounter is dependent on the group velocity and envelope of the THz wave. With envelope of the THz wave represented by $f(t)$, the envelope experienced by the electrons is given by $g(t)=v_gf(t)/(v_e-v_g)$. To realize conversion between the longitudinal and transverse polarizations, time integral of the magnetic field experienced by the electrons must satisfy the condition $\int B(t)dt = \gamma ^2 /(2g_e)$. Here, the field integral is normalized to $B_0 T_0$. Thus, the field integral is directly proportional to $\gamma^2$, thereby leading to an escalation in the required time and magnetic field intensity as the electron energy increases. The spin rotation of high energy electrons can be realized through optimizing the DLW structure to increase $v_g$, and enhancing the pulse energy, as well as broadening the pulse width.

\begin{figure}[t]
	
	\begin{center}
		\flushright
		\includegraphics[width=0.96\linewidth]{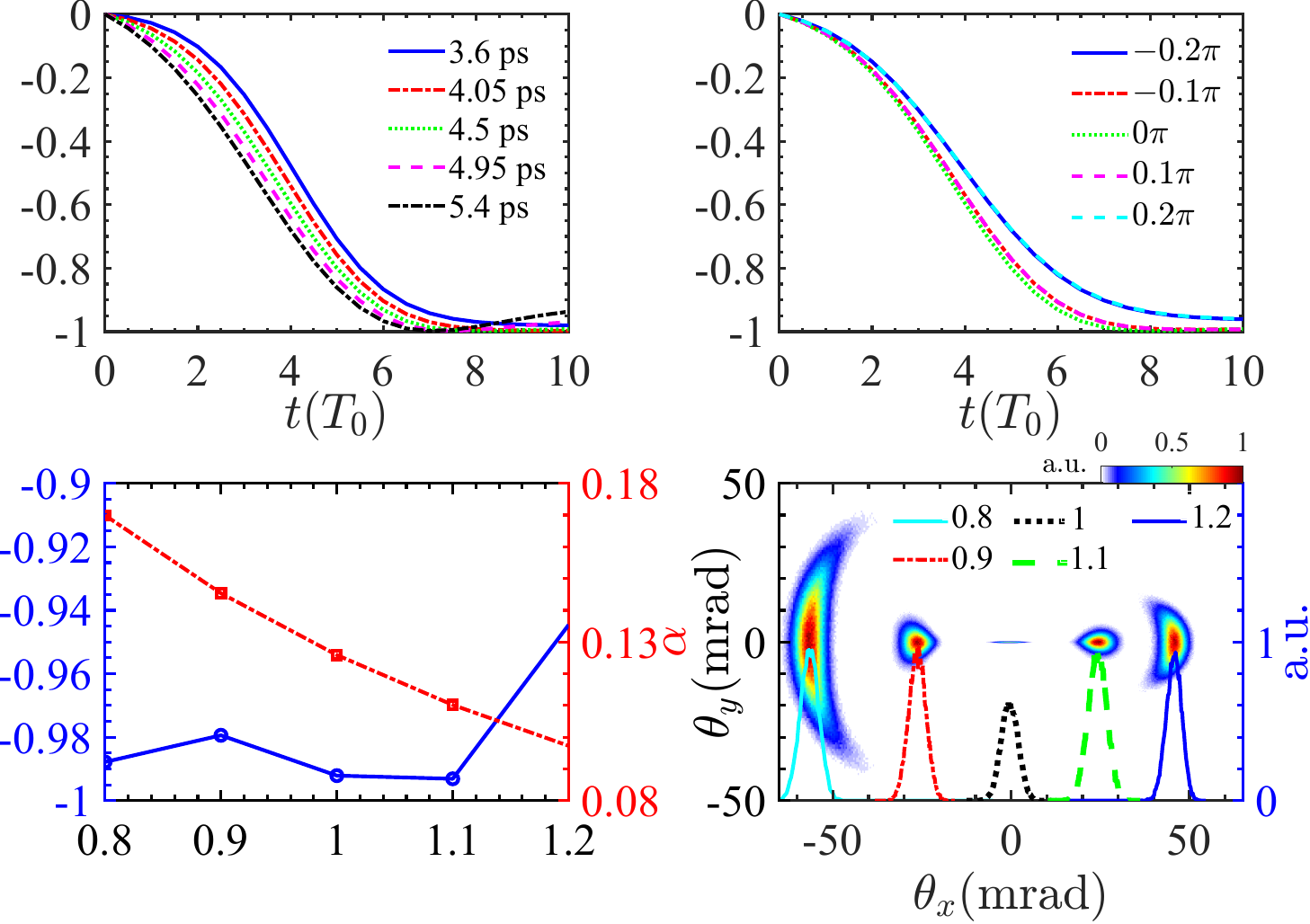}
		\begin{picture}(300,0) 
			\put(30,166){(a)}
			\put(0,142){\rotatebox{90}{$\overline{s_z}$}}
			\put(152,166){(b)}
			\put(123,142){\rotatebox{90}{$\overline{s_z}$}}
			\put(30,82){(c)}
			\put(0,59){\rotatebox{90}{\color{blue}{$\overline{s_z}$}}}
			\put(58,12){ $\varepsilon_e (\varepsilon_0)$}
			\put(152,82){(d)}
			
		\end{picture}
		\setlength{\abovecaptionskip}{-0.7 cm}
		\setlength{\belowcaptionskip}{-0.1 cm}
		\caption{\small{(a) Temporal evolution of average spin-polarization $\overline{s_z}$ for incident THz wave durations of 3.6, 4.05, 4.5, 4.95, and 5.4 ps, respectively. (b) Temporal evolution of $\overline{s_z}$ for THz wave CEP values of $-0.2\pi$, $-0.1\pi$, $0\pi$, $0.1\pi$, and $0.2\pi$, respectively. (c) Dependence of polarization and longitudinal precession frequency coefficient $\alpha$ on the initial electron energy $\varepsilon_e$ when the THz phase velocity is fixed at $0.935c$. Electron energy is normalized to $\varepsilon_0 =0.93 \rm MeV$, which matches the THz phase velocity. (d) Number densities of the electron beams (pcolor) in $(\theta_x,\theta_y)-$space and the corresponding beam profiles (lines) for initial mean electron-beam energies $\varepsilon_e$ equal to $0.8\varepsilon_0$, $0.9\varepsilon_0$, $\varepsilon_0$, $1.1\varepsilon_0$, and $1.2\varepsilon_0$, respectively.} } \label{figvarspeed}
	\end{center}
\end{figure}

As previously mentioned, the $\rm TE_{11}$ mode also serves to compress the energy spread of the electron beam. The transverse Lorentz force $F_{\perp}$ on electrons is positively correlated with $\Delta \beta$~\cite{supp}, where $\Delta\beta = (v_p/c - \beta_e)$ is the difference between normalized electron velocity and THz phase velocity. As depicted in Fig. \ref{fig3}(d) when $\Delta \beta >0$, the force exerted by the transverse magnetic field surpasses that of the electric field. Consequently, the electrons gain momentum along the $x-$axis, leading to deceleration in the longitudinal direction under the influence of the transverse magnetic field. By contrast, when $\Delta \beta <0$, the electrons undergo acceleration in the longitudinal direction. This mechanism contributes to compression of the energy spread.

To assess experimental feasibility, we investigate impact of the THz wave and electron beam parameters on the average degree of polarization of the electron beam, upon its exit from the DLW. Pulse duration of the THz wave exhibits fluctuations and the jitter in its phase can displace electrons from the ideal manipulation phase. These are experimental realities that must be dealt with properly. Results of our investigation of the effects of THz pulse duration fluctuations and phase jitter on spin-polarization, are shown in Figs.~\ref{figvarspeed}(a)-(b). In scenarios featuring a $20\%$ fluctuation in pulse duration, polarization degree of the electron beam remains robustly above $93\%$. For a more conservative case with a $10\%$ variation in pulse duration, the polarization degree demonstrates greater stability (exceeding $97\%$). Our investigations into the effects of phase jitter reveal that, within a fluctuation range of $0.1\pi$, average polarization degree of the electron beam exhibits remarkable insensitivity, sustaining its high value of $99\%$. When the phase jitter extends to a range of $0.2\pi$, the polarization degree of the electron beam is maintained at a commendable level, exceeding $96\%$. This robustness stems from the fact that the optimal manipulation phase aligns with the extremum of the sinusoidal function that characterizes the magnetic field. Within the phase jitter range of $0.1\pi$, the variations in magnetic field intensity experienced by the electrons are negligible, thus preserving the high polarization degree.

To minimize deflection of the electron beam, we align $v_e$ with $v_p$, thereby suppressing $F_{\perp}$. Even if $v_e$ does not perfectly match $v_p$, efficient spin rotation can still be attained despite slight deflection of the electron beam from the DLW's center.
When the electron energy is lower than the energy required to match the THz phase velocity, a high degree of spin-polarization can still be achieved due to a relatively high value of $\alpha$. Similarly, if the electron energy exceeds the energy required for THz matching by $20\%$, resulting from a relatively low value of $\alpha$, it is still possible to obtain an electron beam with a degree of polarization as high as $95\%$, as shown in Fig.~\ref{figvarspeed}(c). In addition, deflection of the electron beam resulting from the transverse Lorentz force is approximately 50 mrad even when the beam is subject to a $20\%$ fluctuation in electron energy, as shown in Fig.~\ref{figvarspeed}(d).

In conclusion, we have put forward a novel flexible method for ultrafast spin rotation of a relativistic lepton beam using a THz wave in a DLW. Our method is feasible with current THz technology and offers a versatile mechanism for spin-manipulation. Flexibility in adjusting EM modes within the DLW adds further potential for spin-manipulation, and holds promise for advancing the development of SP particle beams, which have broad applications in materials science and atomic, nuclear, and high-energy physics, etc.\\


 {\it Acknowledgement:} This work is supported by the National Natural Science Foundation of China (Grants No. U2267204, No. 12275209, No. 12105217), the Foundation of Science and Technology on Plasma Physics Laboratory (No. JCKYS2021212008), Natural Science Basic Research Program of Shaanxi (Grants No. 2023-JC-QN-0091, No. 2024JC-YBQN-0042), the Fundamental Research Funds for Central Universities (No. xzy012023046), and the Shaanxi Fundamental Science Research Project for Mathematics and Physics (Grants No. 22JSY014, No. 22JSQ019, No. 23JSQ006). YIS is supported by an American University of Sharjah Faculty Research Grant (FRG24-E-S29) and acknowledges hospitality at the School of Physics, Xi'an Jiaotong University.


\bibliography{refs}

\end{document}